# Sputtered Aluminum Nitride Waveguides for the Telecommunication Spectrum with less than 0.16 dB/cm Loss

Radhakant Singh,[1,2] Mohit Raghuwanshi,[3] Balasubramanian Sundarapandian,[3] Rijil Thomas,[1] Lutz Kirste,[3] Stephan Suckow,[1,*] Max Lemme[1,2,*]

[1]*AMO GmbH, Advanced Microelectronic Center Aachen, 52074, Germany*
[2]*RWTH Aachen University, Chair of Electronic Devices, 52074 Aachen, Germany*
[3]*Fraunhofer Institute for Applied Solid State Physics IAF, 79108 Freiburg im Breisgau, Germany*
**Email: suckow@amo.de, lemme@amo.de*

**Abstract:** We report the fabrication and characterization of photonic waveguides from sputtered aluminum nitride (AlN). The AlN films were deposited on 6" silicon substrates with a 3 µm buried silicon oxide layer using reactive DC magnetron sputtering at a temperature of 700°C. The resulting uncladded polycrystalline waveguides exhibit propagation losses of 0.137 ± 0.005 dB/cm at wavelengths of 1310 nm and 0.154 ± 0.008 dB/cm at a wavelength of 1550 nm in the TE polarization. These results are the best reported for sputtered AlN waveguides in the C-band and the first report in the O-band. These performances are comparable to those of the best-reported AlN waveguides, which are epitaxially grown by metal-organic chemical vapor deposition (MOCVD) on sapphire substrates. Our findings highlight the potential of sputtered AlN for photonic platforms working in the telecom spectrum.

## 1. Introduction

Aluminum nitride (AlN) is an excellent candidate material for photonic waveguide (WG) applications because it has a large band gap of 6.2 eV and, therefore, a wide transparency window from ultraviolet (UV) to mid-infrared wavelengths **[1,2]**. The AlN platform has strengths similar to those of the established silicon nitride (SiN) platform, such as moderate optical confinement, allowing for similarly low-loss waveguiding in the telecom bands. The advantages of AlN over SiN are (i) an enhanced transparency window extending to the blue and near-UV spectra, (ii) the ability to handle higher optical power due to less two-photon absorption **[3]**, (iii) significantly higher thermal conductivity leading to better heat dissipation **[4,5]**, and (iv) a noncentrosymmetric crystal structure, giving rise to remarkable $2^{nd}$-order nonlinearity in addition to $3^{rd}$-order optical nonlinearity. Therefore, it can be used for electro-optic modulation **[6]** and $2^{nd}$ or $3^{rd}$ harmonic generation **[7,8]**. AlN is also used in piezoelectric applications, rendering it suitable for use in electromechanical and optomechanical devices **[9,10]**. Furthermore, the AlN fabrication process is compatible with the complementary metal–oxide–semiconductor (CMOS) fabrication process, facilitating scalable wafer-level production of cost-effective and highly reliable AlN-based integrated photonics devices, both active and passive **[3,11]**.

AlN grown on sapphire substrates by metal–organic vapor phase epitaxy (MOVPE), also known as metal–organic chemical vapor deposition (MOCVD), at a process temperature of approximately 1200°C achieved propagation losses of 0.14 dB/cm and 0.2 dB/cm for the TE and TM modes, respectively **[12]**. However, sputtering is favored over MOCVD because of its compatibility with substrates other than sapphire, the absence of organic contamination, and the ability to deposit films at lower temperatures. AlN WGs prepared via sputter deposition have been used in the past, but the presence of impurities, a polycrystalline structure, and greater optical losses have been reported. Here, we investigate uncladded sputtered AlN WGs made from highly optimized AlN films. We achieved the lowest waveguide losses reported for sputtered AlN waveguides in both the telecommunication O- and C-bands.

## 2. Fabrication

3 μm thick silicon dioxide ($SiO_2$) was thermally grown in a Centrotherm furnace on 6” Si (001) wafers to isolate the WGs from the Si substrate optically. The AlN films were deposited via an Evatec Clusterline 200II planar magnetron sputtering module. A 304 mm diameter Al target with a purity of 99.9995% was used to deposit AlN films on the thermally grown $SiO_2$. In our previous works **[13,14]**, we reported that the growth temperature and process gas are crucial for controlling the crystallographic texture and optical properties of AlN and reducing impurities in the films. Therefore, AlN was sputtered at 700°C in a pure nitrogen atmosphere to achieve high-quality AlN films with low impurity levels. We used a power of 5500 W, a gas flow of 40 sccm, and a target-to-substrate distance of 10 mm. Additionally, a base pressure ranging between $8 \times 10^{-8}$ and $1\times10^{-7}$ mbar was maintained to minimize the amount of oxygen and other impurities in the sputter chamber.

Patterning the AlN WGs required a thin XHRiC (Brewer Science Ltd.) i-line bottom antireflection coating (BARC) to ensure optimal feature size control and lithographic performance. The positive tone photoresist AZmir-701 (Microchemicals GmbH) was then deposited on the BARC-coated wafer via an EVG-150 automated resist spin-coater. The WGs were then defined via optical projection lithography using a Canon-FPA 3000 i5+ i-line stepper and patterned via reactive ion etching (RIE) using an Oxford Instruments PlasmaLab System 100 tool with chlorine-based chemistry ($BCl_3$, $Cl_2$, and He). The fabrication process flow is shown in Figure 1a.

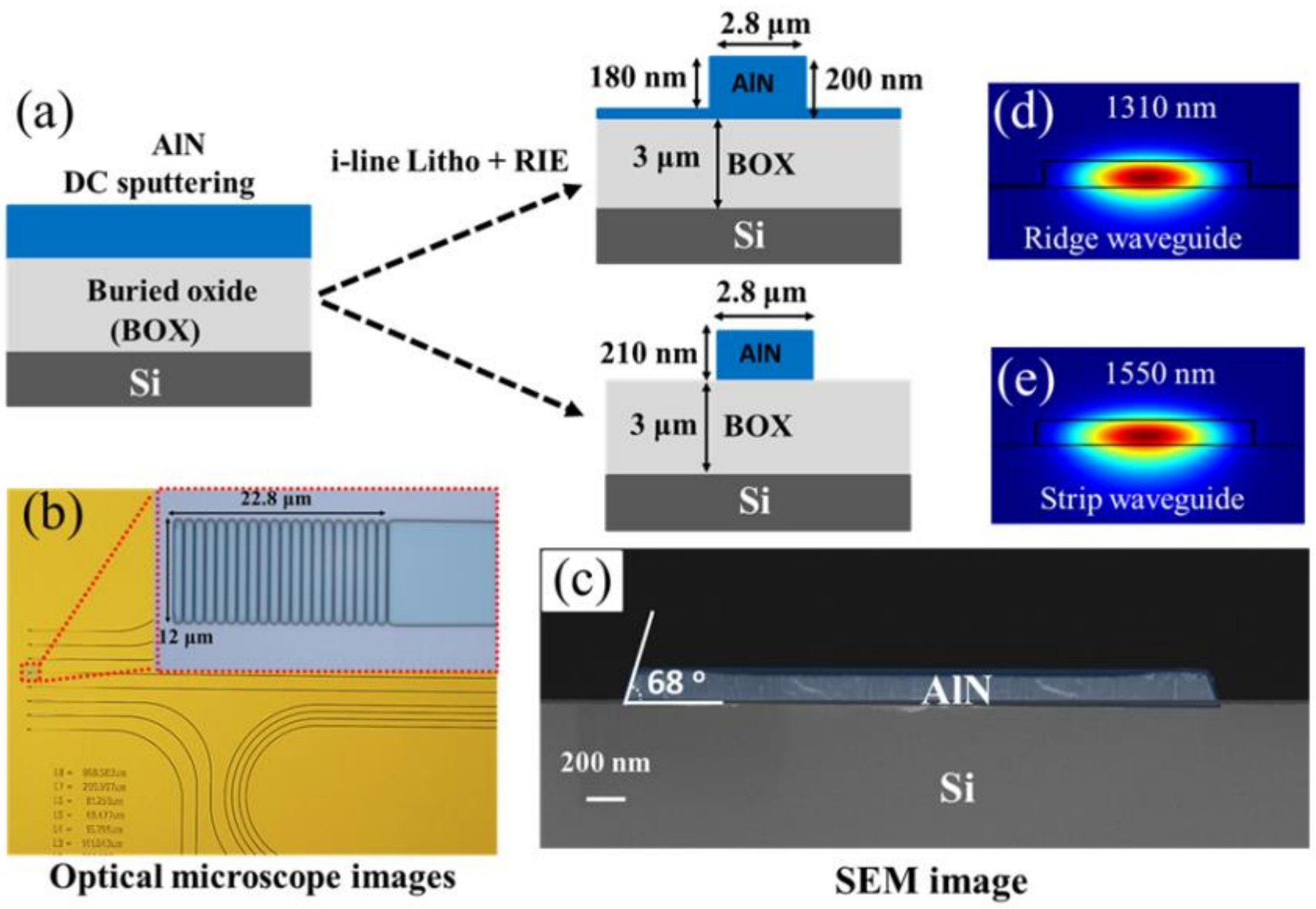

Figure 1: (a) AlN waveguide fabrication process flow, resulting in two different waveguide cross sections. (b) Optical image of the fabricated grating coupler and waveguides. (c) Representative cross-sectional SEM image. (d) and (e) represent the mode profile of the $TE_0$ mode for the Ridge waveguide at 1310 nm and the Strip waveguide at 1550 nm, respectively.

We fabricated ridge WGs with a width of 2.8 µm, a thickness of 200 nm, and an etch depth of 180 nm for the O-band and strip WGs with a width of 2.8 µm and a thickness of 210 nm for the C-band (see schematic in Figure 1a). Figure 1b shows an optical microscope image of the fabricated WGs. The sidewall angle of a representative test AlN WG with a height of ~110 nm on a Si substrate is 68° after RIE (Figure 1c). Figures 1d and 1e show the mode profiles of the $TE_0$ modes for the ridge WG at 1310 nm and the strip WG at 1550 nm, respectively. Without $SiO_2$ cladding, the ridge WG supports three modes at 1310 nm ($TE_0$, $TE_1$ and weakly guiding $TE_2$), and the strip WG supports two modes at 1550 nm ($TE_0$ and weakly guiding $TE_1$).

## 3. Results and Discussions

The AlN films were first characterized unpatterned. The refractive indices and extinction coefficients of the films were measured using a J. A Woollam spectroscopic ellipsometer. (Figure 2a) The real part of the refractive index is comparable to that reported by others **[15]**. The determined extinction coefficient is less than $10^{-11}$ over the full telecommunication range, which signals negligible material absorption loss and may be below the detection limit of the tool. A Bruker ICON atomic force microscope (AFM) was used to examine the surface morphology (Figure 2b). The root mean square (RMS) surface roughness ($R_q$) of the films is 2.1 nm, and the average grain diameter ($d_{grain}$) is 31 nm. X-ray diffraction (XRD) analysis was performed with a PANalytical X'Pert Pro MRD diffractometer equipped with a Ge-220 hybrid monochromator providing Cu-Kα1 radiation. Figure 2c shows the XRD 2θ/θ scans of wurtzite-type AlN sputtered on $SiO_2$/Si. The XRD pattern shows that even though AlN is deposited on amorphous $SiO_2$, it has a strong texture along the [0001] direction and a ω-FWHM of approximately 3° for the 0002 reflection (Figure 2d).

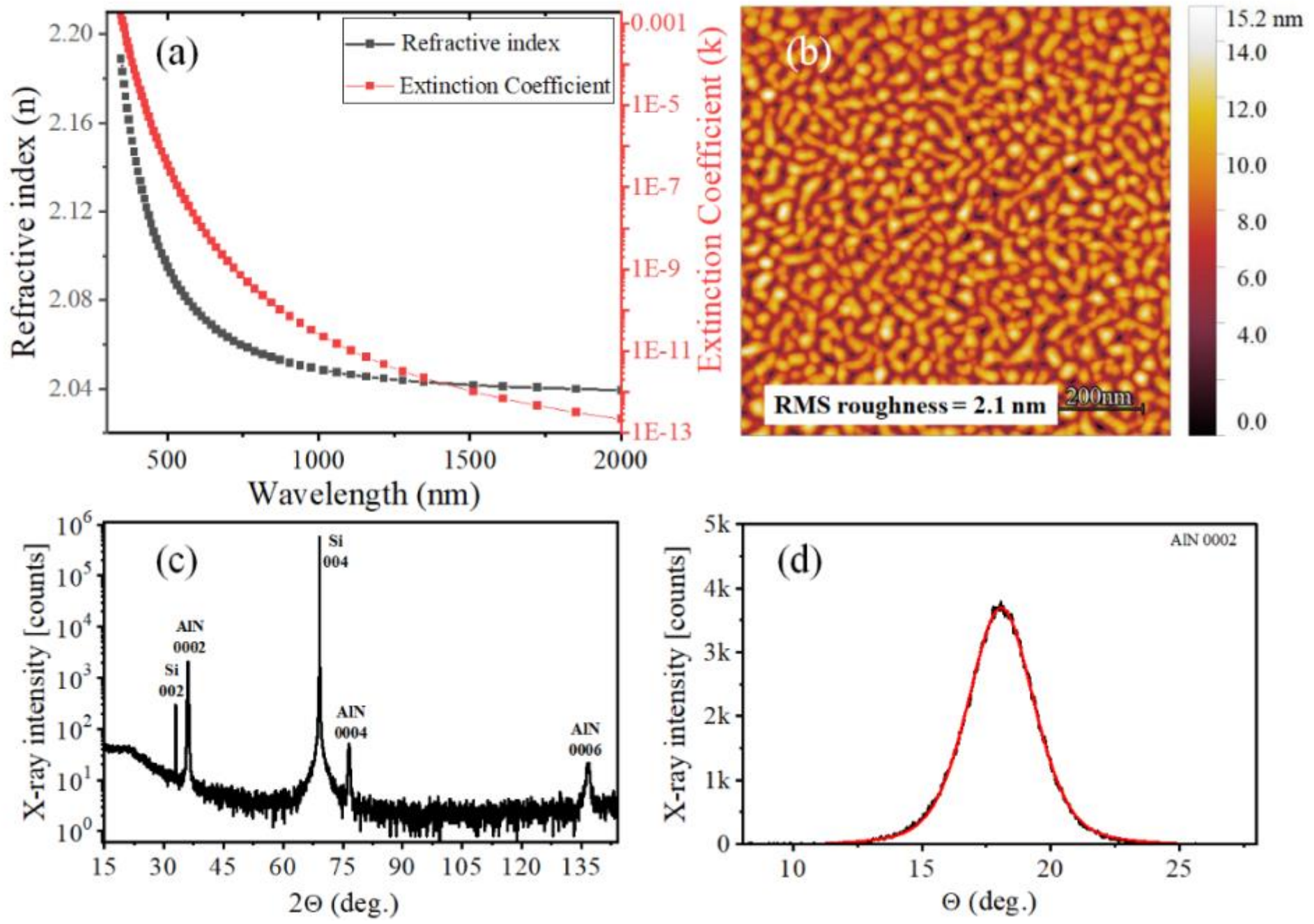

Figure 2: (a) Refractive index (n) and extinction coefficient (k) of our AlN film on a 6" Si substrate versus wavelength and (b) AFM micrograph of the deposited AlN films on a 6" oxide wafer. The root-mean-square (RMS) roughness is 1.7 nm. (c) XRD 2Θ/Θ pattern of our AlN deposited at 700°C on $SiO_2$ and (d) ω-scan of the AlN 0002 reflection showing an ω-FWHM of approximately 3°.

The propagation losses of the uncladded optical WGs were characterized with the cutback method. A polarization-maintaining (PM), single-mode (SM) fiber connected to a broadband tunable laser (Agilent 8164A) was used to couple the fundamental TE mode into the AlN WGs through grating couplers. The laser input power was maintained at 1 mW, while the input wavelength was swept from 1260 nm to 1370 nm for the O-band and from 1520 nm to 1620 nm for the C-band in steps of 50 pm for all measurements. The output optical transmission spectra were collected via a multimode fiber (MMF) connected to a power meter (Agilent 81635A). All the in-coupling and out-coupling measurements were performed at an angle of 15° to the normal.

The mask design included eight WGs with lengths ranging from 1.5 cm to 109 cm and a minimum bend radius of 500 µm. The propagation losses were determined via the cutback method by plotting the transmission through these WGs (see spectra in Figure 3a & d for the O- and C-bands, respectively) versus their lengths (see Figure 3b & e for the O- and C-bands, respectively) and performing a linear fit to the data. The slope of the fit yields propagation loss, and the y-intercept contains all losses independent of the WG length. The y-intercept mainly comprises the grating coupler losses for in-coupling and out-coupling. A good linear fit quality signals trustworthy data and negligible variations in the y-intercept between different WGs.

Figure 3c shows the average propagation loss of all eight dies measured in the O-band, weighted by the uncertainties of the linear fits to the cutback plots of all dies. The data range from 0.12 to 0.15 dB/cm over the entire spectrum, with an average WG loss of $(0.137 \pm 0.005)$ dB/cm at 1310 nm. Figure 3f shows the weighted average propagation loss of all three dies measured, and the individual results in the C-band, ranging from 0.14 to 0.2 dB/cm over the entire spectrum. The average WG loss is $(0.154 \pm 0.008)$ dB/cm at 1550 nm. The error margins at wavelengths of 1310 nm and 1550 nm are the standard deviations of the loss data from all the measured dies. Note that all the measured dies are included in the analysis. However, we have removed data from defective WGs, which can easily occur due to, e.g., dust particles on our long, uncladded WGs. For the O-band analysis, we used 48 out of 64 waveguides (75% yield), and for the C-band analysis, we included 18 out of 24 WGs (75% yield).

The increasing loss toward shorter wavelengths in the O-band may be attributed to increased scattering from the surface roughness. For the C-band, we attribute the increasing loss toward longer wavelengths to substrate leakage, which could be avoided with thicker $SiO_2$ below AlN or with thicker AlN layers for tighter mode confinement. The coupler losses are ~12 dB/coupler for both 1550 nm and 1310 nm, as they were not optimized for our AlN layer.

Table 1 shows a comparison of the state-of-the-art of AlN WG losses compared with our work.

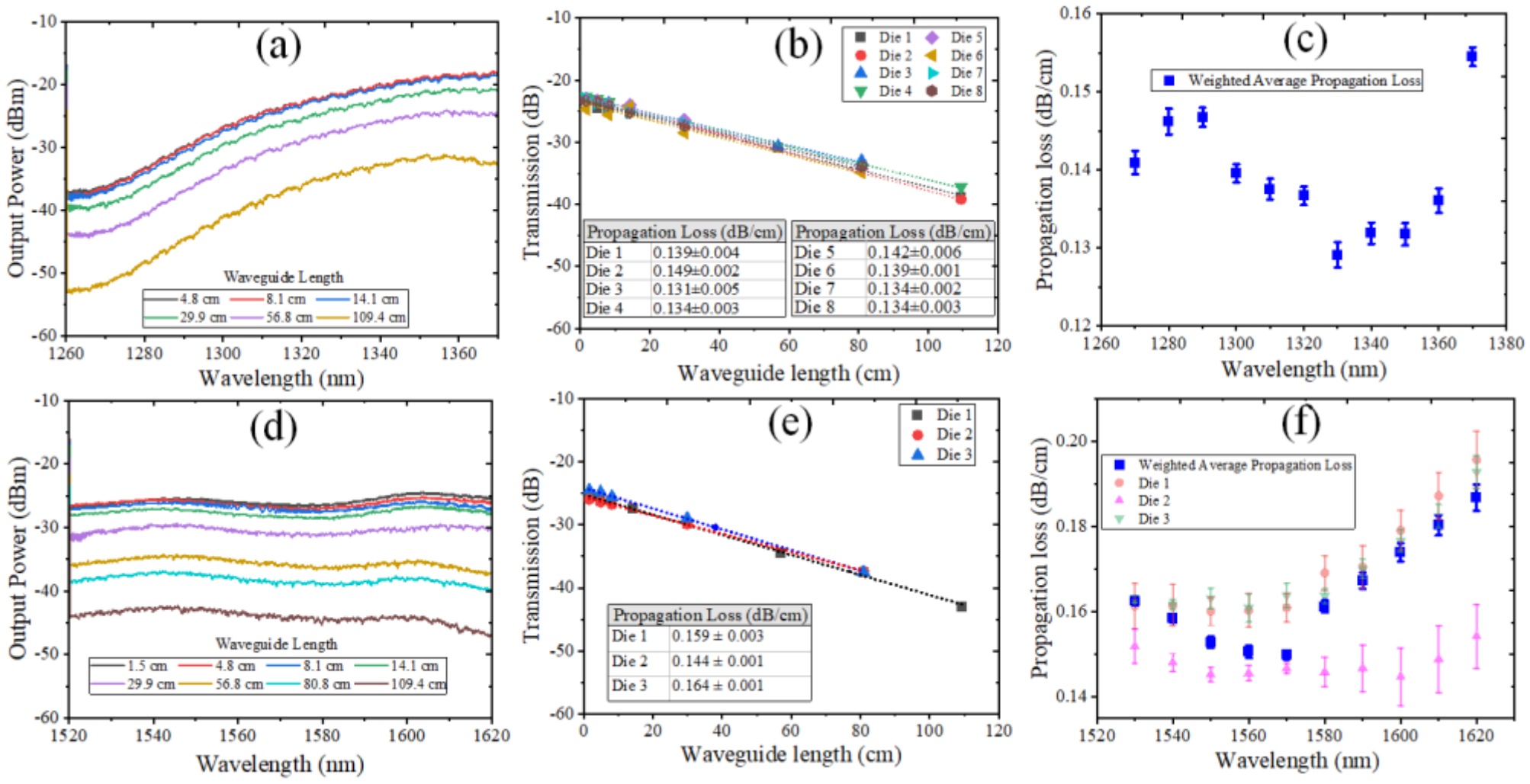

Figure 3: (a & d) Experimentally measured AlN waveguide transmission spectra in the O-band (a) and C-band (d). Cutback plots in the O-band (b) and the C-band (e). In both cases, “die 1” is shown in (a) and (d), respectively. Propagation losses in the full spectra measured around the O-band (c) and C-band (f).

Table 1: Optical waveguide loss of AlN waveguides in the O- and C-bands.

| **Reference/ Year** | **Wave- length** | **Device Structure** | **AlN Fabrication Technique** | **Waveguide Geometry/ width x height** | **Optical Loss** |
|---|---|---|---|---|---|
| **[12]**/2017 | 1554 n m | Microring resonator ($SiO_2$ cladding) | MOCVD AlN on Sapphire (1200°C) | Strip waveguide 3.5 µm x 1.2 µm | 0.14 dB/cm ($TE_0$) 0.2 dB/cm ($TM_0$) |
| **[16]**/2020 | 1550 n m | Microring resonator ($SiO_2$ cladding) | Magnetron reactive sputtering (800°C cyclic RTA) | Rib waveguide (etch depth 400 nm)/ 600 nm x 1 µm | 0.76 dB/cm ($TM_0$) ~0.87 dB/cm ($TE_0$) |
| This work/2024 | 1550 n m | Waveguide (Air cladding) | Magnetron reactive sputtering (700°C) | Strip waveguide/ 2.8 µm x 210 nm | 0.155 dB/cm ($TE_0$) |
| This work/2024 | 1310 n m | Waveguide (Air cladding) | Magnetron reactive sputtering (700°C) | Rib waveguide (etch depth 180 nm)/ 2.8 µm x 200 nm | 0.137 dB/cm ($TE_0$) |

## 4. Conclusion

In summary, AlN films were sputter deposited on $SiO_2$ at 700°C at a low background pressure to minimize the oxygen content and to maximize the quality of the film. We fabricated uncladded AlN WGs with grating couplers and measured their optical losses. We achieved optical losses of 0.137 $\pm$ 0.005 dB/cm and 0.154 $\pm$ 0.008 dB/cm at wavelengths of 1310 nm and 1550 nm, respectively. The data are based on 2 wafers and 88 measured WGs, of which 75% were considered defect-free and included in the analysis. This optical performance matches the previously reported best value achieved with epitaxially grown AlN on sapphire via MOCVD. Moreover, adding silicon dioxide cladding and/or annealing is expected to further improve the optical losses achievable with our AlN.

**Funding:** This work received funding from BMBF under grant agreement No. 402105 (ATIQ).